\documentclass[twocolumn,superscriptaddress,showpacs,pra]{revtex4}
\usepackage{epsfig}
\usepackage{epstopdf}
\usepackage{color}
\usepackage{delarray}
\usepackage{amsmath, amssymb}
\usepackage{bm}
\usepackage{float}
\usepackage{graphicx}
\usepackage{color}

\begin{document}
\title{Freezing Optical Rogue Waves by Zeno Dynamics}

\author{Cihan Bay\i nd\i r}\email{cihanbayindir@gmail.com}
\affiliation{Faculty of Engineering, Isik University, Sile, Istanbul, 34980, Turkey}

\author{Fatih Ozaydin}
\affiliation{Department of Information Technologies, Isik University, Sile, Istanbul, 34980, Turkey}


\begin{abstract}

We investigate the Zeno dynamics of the optical rogue waves. Considering their usage in modeling rogue wave dynamics, we analyze the Zeno dynamics of the Akhmediev breathers, Peregrine and Akhmediev-Peregrine soliton solutions of the nonlinear Schr\"{o}dinger equation. We show that frequent measurements of the wave inhibits its movement in the observation domain for each of these solutions. We analyze the spectra of the rogue waves under Zeno dynamics. We also analyze the effect of observation frequency on the rogue wave profile and on the probability of lingering of the wave in the observation domain. Our results can find potential applications in optics including nonlinear phenomena.


\end{abstract}
\maketitle

\section{Introduction}

Rogue (freak) waves can be described as high amplitude waves with a height bigger than $2-2.2$ times the significant waveheight in a wavefield. Their studies have become extensive in recent years \cite{Akhmediev2009b, Akhmediev2009a, bayindir2016, Akhmediev2011}. The research has emerged with the investigation of one of the simplest nonlinear models, which is the nonlinear Schr\"{o}dinger equation (NLSE)\cite{Akhmediev2009b}. Discovery of the unexpected rogue wave solutions of the NLSE resulted in seminal studies of rogue wave dynamics, such as in Ref.\cite{Akhmediev2009b}. Their existence is not necessarily restricted to optical media \cite{FirstOpticalRW}, they can also be observed in hydrodynamics, Bose-Einstein condensation, acoustics and finance, just to name a few \cite{Akhmediev2009b, bayindir2016}. It is natural to expect that in a medium whose dynamics are described by the NLSE and NLSE like equations, rogue waves can also emerge.
In this study we consider optical rogue waves for which analyzing the dynamics, shapes and statistics of rogue wavy optical fields are crucially important to satisfy certain power and communication constraints.

On the other hand, quantum Zeno dynamics \cite{MisraSudarshan,Facchi2008JPA}, which is the inhibition of the evolution of an unstable quantum state by appropriate frequent observations during a time interval has attracted an intense attention in quantum science, usually for protecting the quantum system from decaying due to inevitable interactions with its environment. It emerged that the observation alters the evolution of an atomic particle, even can stop it \cite{HarochePRL2010, HarochePRA2012, HarocheNP2014}. Duan and Guo showed that the dissipation of two particles can be prevented \cite{Guo1998PRL,Guo1998PRA}, Viola and Lloyd proposed a dynamical suppression of decoherence of qubit systems \cite{Lloyd1998PRA}, Maniscalco et al. proposed a strategy to fight against the decoherence of the entanglement of two atoms in a lossy resonator \cite{Maniscalco2008PRL}, Nourmandipour et al. studied Zeno and anti-Zeno effects on the entanglement dynamics of dissipative qubits coupled to a common bath \cite{Nourmandipour2016JOSAB}, and Bernu et al. froze the coherent field growth in a cavity \cite{Bernu2008PRL}.
Very recently, Facchi et al. studied the large-time limit of the quantum Zeno effect \cite{Facchi17JMP}.

Quantum Zeno dynamics can also be used for realizing controlled operations and creating entanglement. Creation of entanglement is a major issue in quantum information science, requiring controlled operations such as CNOT gates between qubits, which is usually a demanding task. As the number of qubits exceeds two, multipartite entanglement emerges in inequivalent classes such as GHZ, W and cluster states, which cannot be transformed into each other via local operations and classical communications. The preparation of multipartite entangled states -especially W states- require not only even more controlled operations but also novel methods \cite{OzaydinW1, OzaydinW2, OzaydinW3, OzaydinW4, OzaydinW5}. Wang et al. proposed a collective threshold measurement scheme for creating bipartite entanglement, avoiding the difficulty of applying CNOT gates or performing Bell measurements \cite{Wang2008PRA}, which can be extended to multipartite entangled states. Chen et al. proposed to use Zeno dynamics for generation of W states robust against decoherence and photon loss \cite{Chen2016OptComm} and Barontini et al. experimentally demonstrated the deterministic generation of W states by quantum Zeno dynamics \cite{Barontini2015Science}. Nakazoto et al. further showed that purifying quantum systems is possible via Zeno-like measurements \cite{Nakazoto20013PRL}.

Optical analogue of the quantum Zeno effect has been receiving an increasing attention.
Yamane et al. reported Zeno effect in optical fibers \cite{Yamane2001OptComm}.
Longhi proposed an optical lattice model including tunneling-coupled waveguides for the observation of the optical Zeno effect \cite{Longhi2006PRL}.
Leung and Ralph proposed a distillation method for improving the fidelity of optical Zeno gates \cite{Leung2006PRA}.
Biagioni et al. experimentally demonstrated the optical Zeno effect by scanning tunneling optical microscopy \cite{Biagioni2008OptExp}.
Abdullaev et al. showed that it is possible to observe the optical analog of not only linear but also nonlinear quantum Zeno effects in a simple coupler and they further proposed a setup for the experimental demonstration of these effects \cite{Abdullaev2011PRA}.
McCusker et al. utilized quantum Zeno effect for the experimental demonstration of the interaction-free all-optical switching \cite{Kumar2013PRL}.
Thapliyal et al. studied quantum Zeno and anti-Zeno effects in nonlinear optical couplers \cite{Thapliyal2016PRA}.

In this paper we numerically investigate the optical analogue of quantum Zeno dynamics of the rogue waves that are encountered in optics. With this motivation, in the second section of this paper we review the NLSE and the split-step Fourier method for its numerical solution. We also review a procedure applied to wavefunction to model the Zeno dynamics of an observed system. In the third section of this paper, we analyze the  Zeno dynamics of the Akhmediev breathers, Peregrine and Akhmediev-Peregrine soliton solutions of the NLSE, which are used as models to describe the rogue waves. We show that frequent measurements of the wave inhibits the movement of the wave in the observation domain for each of these types of rogue waves. We also analyze the spectra of the rogue waves under Zeno dynamics and discuss the effect of observation frequency on the rogue wave profile and on the probability of lingering of the wave in the observation domain. In the last section we conclude our work and summarize future research tasks.

\section{Nonlinear Schr\"{o}dinger Equation and Zeno Effect}

It was shown that all the features of linear quantum mechanics can be reproduced by NLSE \cite{Richardson2014PRA}, and quantum NLSE can accurately describe quantum optical solitons in photonic waveguides with Kerr nonlinearity \cite{Carter1987PRL,Drummond1987JOSAB,Lai1989PRA844,Lai1989PRA854}.
The bosonic matter wave field for weakly interacting ultracold atoms in a Bose-Einstein condensate, evolves according to quantum NLSE \cite{LeggettRMP2001,MorschRMP2006}.
Many nonlinear phenomena observed in fiber optics are generally studied in the frame of the NLSE \cite{Akhmediev2009b}. Optical rogue waves are one of those phenomena and rational rogue wave soliton solutions of the NLSE are accepted as accurate optical rogue wave models \cite{Akhmediev2009b}. In order the analyze the Zeno dynamics of rogue waves, we consider the nondimensional NLSE given as
\begin{equation}
i\psi_t + \frac{1}{2} \psi_{xx} +  \left|\psi \right|^2 \psi =0,
\label{eq01}
\end{equation}
where $x$ and $t$ are the spatial and temporal variables, respectively, $i$ is the imaginary number, and $\psi$ is the complex amplitude. It is known that the NLSE given by Eq.(\ref{eq01}) admits many different types of analytical solutions. Some of these solutions are reviewed in the next section of this paper. For arbitrary wave profiles, where the analytical solution is unknown, the NLSE can be numerically solved by a split-step Fourier method (SSFM), which is one of the most commonly used forms of the spectral methods. Similar to other spectral methods, the spatial derivatives are calculated using spectral techniques in SSFM. Some applications of the spectral techniques can be seen in Refs.\cite{bay2009, Bay_arxNoisyTun, BayTWMS2016, bay_cssfm, Agrawal, Bay_arxNoisyTunKEE, demiray, Bay_arxEarlyDetectCS, Karjadi2010, Karjadi2012, Bay_cssfmarx, bayindir2016nature, Bay_arxChaotCurNLS, BayPRE1, BayPRE2, Bay_CSRM} and their more comprehensive analysis can be seen in Refs.\cite{canuto, trefethen}.

 The temporal derivatives in the governing equations is calculated using time integration schemes such as Adams-Bashforth and Runge-Kutta, etc. \cite{canuto, trefethen, demiray}. However, SSFM uses an exponential time stepping function for this purpose. SSFM is based on the idea of splitting the equation into two parts, namely the linear and the nonlinear parts. Then time stepping is performed starting from the initial conditions. In a possible splitting we take the first part of the NLSE as
\begin{equation}
i\psi_t= -\left| \psi \right|^2\psi
\label{eq02}
\end{equation}
which can exactly be solved as
\begin{equation}
\tilde{\psi}(x,t_0+\Delta t)=e^{i \left| \psi(x,t_0)\right|^2  \Delta t}\ \psi_0,
\label{eq03}
\end{equation}
where $\Delta t$ is the time step and $\psi_0=\psi(x,t_0)$ is the initial condition. The second part of the NLSE can be written as
\begin{equation}
i\psi_t=- \frac{1}{2} \psi_{xx}.
\label{eq04}
\end{equation}
Using a Fourier series expansion we obtain
 \begin{equation}
\psi(x,t_0+\Delta t)=F^{-1} \left[e^{-i k^2 /2 \Delta t}F[\tilde{\psi}(x,t_0+\Delta t) ] \right],
\label{eq05}
\end{equation}
where $k$ is the wavenumber \cite{bay_cssfm, Agrawal}. Substituting Eq.(\ref{eq03}) into Eq.(\ref{eq05}), the final form of the SSFM becomes
 \begin{equation}
\psi(x,t_0+\Delta t)=F^{-1} \left[e^{-i k^2 /2\Delta t}F[ e^{i \left| \psi(x,t_0)\right|^2 \Delta t}\ \psi_0 ] \right].
\label{eq06}
\end{equation}
Starting from the initial conditions, the time integration of the NLSE can be done by the SSFM. Two fast Fourier transform (FFT) operations per time step are needed for this form of the SSFM. The time step is selected as $\Delta t=10^{-3}$, which does not cause a stability problem. The number of spectral components are taken as $M=2048$ in order to use the FFT routines efficiently.

Although it is known that the decay of an atomic particle can be inhibited by Zeno dynamics, it remains an open question whether the rogue waves in the quantized optical fields in the frame of the NLSE can be stopped by Zeno dynamics. In this paper we analyze the Zeno dynamics of such rogue waves by using the SSFM reviewed above. Although analytical solution of the NLSE is known and used as initial conditions in time stepping of SSFM, after a positive Zeno measurement the wavefunction becomes complicated thus numerical solution is needed. 
Recently a theoretical wavefunction formulation of the quantum Zeno dynamics is proposed in Ref.\cite{PorrasFreeZeno}, used in Refs.\cite{Porras_Zenotunnelmimick, Porras_diffractionspread} and experimentally tested in Ref.\cite{Porras_zeno_clss_opt}. In this formulation, after a positive measurement the particle is found in the observation domain of $[-L,L]$ with a wavefunction of $\psi_T(x,t)=\psi(x,t)\textnormal{rect}(x/L)/\sqrt{P}$ where $P=\int_{-L}^L \left| \psi \left(x,t \right) \right|^2 dx$, and $\textnormal{rect}(x/L)=1$ for $-L \leq x \leq L$, and $0$ elsewhere \cite{PorrasFreeZeno}. Between two successive positive measurements, the wave evolves according to NLSE. This cycle can be summarized as
\begin{equation}
\begin{split}
\psi_T &  \left(x,\frac{(n-1)t}{N} \right) \stackrel{evolve}{\rightarrow} \psi \left(x,\frac{nt}{N} \right) \stackrel{measure}{\rightarrow} \\
&    \psi_T \left(x,\frac{nt}{N} \right)= \psi \left(x,\frac{nt}{N} \right) \frac{\textnormal{rect(x/L)}}{\sqrt{P_N^{n}}}
\label{eq07}
\end{split}
\end{equation}
where $n$ is the observation index, $N$ is the number of observations \cite{PorrasFreeZeno}, and
 \begin{equation}
P_N^{n}=\int_{-L}^L \left| \psi \left(x,\frac{nt}{N} \right) \right|^2 dx.
\label{eq08}
\end{equation}
The cumulative probability of finding the wave in the observation domain becomes \cite{PorrasFreeZeno}
 \begin{equation}
P_N=\prod_{n=1}^N P_N^{n}.
\label{eq09}
\end{equation}
Using the momentum representation of the linear Schr\"{o}dinger equation and analogy of optical wave dynamics of Fabry-Perot resonator, an analytical derivation of the lingering probability of an atomic particle in the interval of $[-L,L]$ after $n^{th}$ measurement is given as
 \begin{equation}
P_N^{n}\approx 1-0.12 \left(\frac{4}{\pi} \right)^2  \left(\frac{2 \pi t}{N} \right)^{3/2}
\label{eq10}
\end{equation}
in \cite{PorrasFreeZeno}. After $N$ measurements the cumulative probability of finding the particle in the observation domain becomes
 \begin{equation}
P_N \approx \left(1-0.12 \left(\frac{4}{\pi} \right)^2  \left(\frac{2 \pi t}{N} \right)^{3/2} \right)^N
\label{eq11}
\end{equation}
which can further be simplified using Newton's binomial theorem \cite{PorrasFreeZeno}. The reader is referred to Ref.\cite{PorrasFreeZeno} for the details of the derivation of these relations. We compare the analytical relations given in Eqs.(\ref{eq10})-(\ref{eq11}) with the numerical probability calculations in the next section of this paper.

\section{Results and Discussion}

\subsection{Freezing Akhmediev Breathers}
In order to study the Zeno dynamics of rogue waves we first consider the Akhmediev breather (AB) solution of the NLSE given in Eq.(\ref{eq01}). It is known that the NLSE admits a solution in the form of
\begin{equation}
\psi_{AB}=\left[1+\frac{2(1-2a) \cosh{(bt)}+ i b \sinh{(bt)}}{\sqrt{2a} \cos{(\lambda x)}-\cosh{(bt)} }  \right] \exp{[it]}
\label{eq12}
\end{equation}
where $a$ is a free parameter, $\lambda=2 \sqrt{1-2a}$ and $b=\sqrt{8a(1-2a)}$ \ \cite{AkhmedievBreather, AkhmedievBreatherExp}. This solution is known as AB and plays an essential role in describing the modulation instability mechanism and rogue wave generation. Experiments have demonstrated that the ABs can exist in optical fibers \cite{AkhmedievBreatherExp}. The triangularization of the Fourier spectrum, i.e. the triangular supercontinuum generation, plays a key role in the generation and early detection mechanisms of rogue waves. Thus, we also examine the spectra of ABs under Zeno effect.

\begin{figure}[ht!]
\begin{center}
   \includegraphics[width=3.4in]{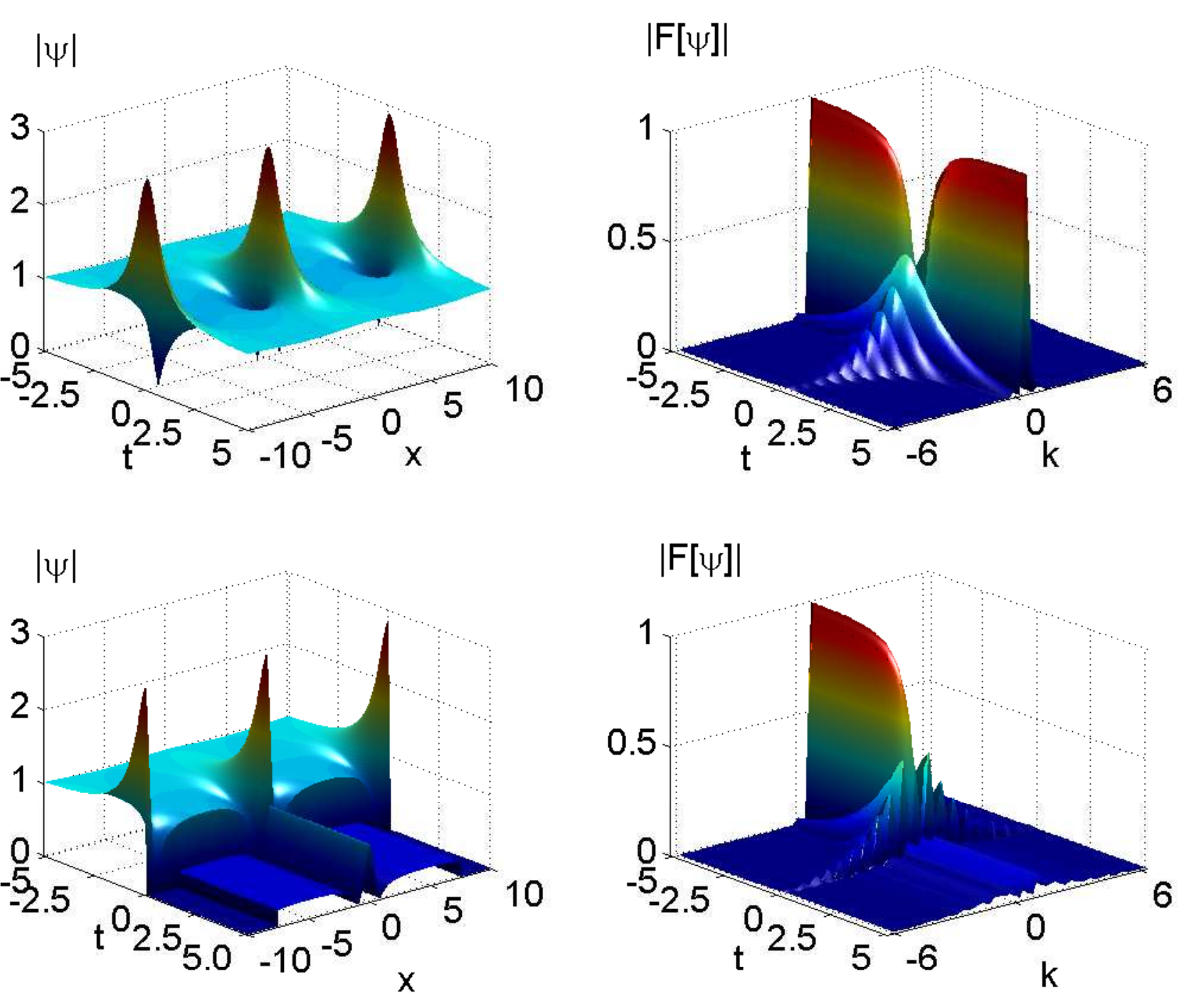}
  \end{center}
\caption{\small Zeno dynamics of an Akhmediev breather with $a=0.45$ a) freely evolving wave b) its Fourier spectrum c) continuously observed Akhmediev breather in $[-7.5,7.5]$ during $t=[0-5]$  d) its Fourier spectrum.}
  \label{fig1}
\end{figure}

In Fig.~\ref{fig1}, 3D plots of the AB for $a=0.45$ and its Fourier spectrum are depicted in the first row. This AB is freely evolving for the time interval of $t=[-5,5]$. In the second row of Fig.~\ref{fig1}, we present the AB under Zeno effect and its Fourier spectrum. This AB evolved freely  in the time interval of $t=[-5,0]$ and it was continuously subjected to Zeno observations in the time interval of $t=[0,5]$ within $L=[-7.5,7.5]$. For a better visualization of the effect of Zeno observations, the wave profile is not normalized in this figure.

\begin{figure}[htb!]
\begin{center}
   \includegraphics[width=3.4in]{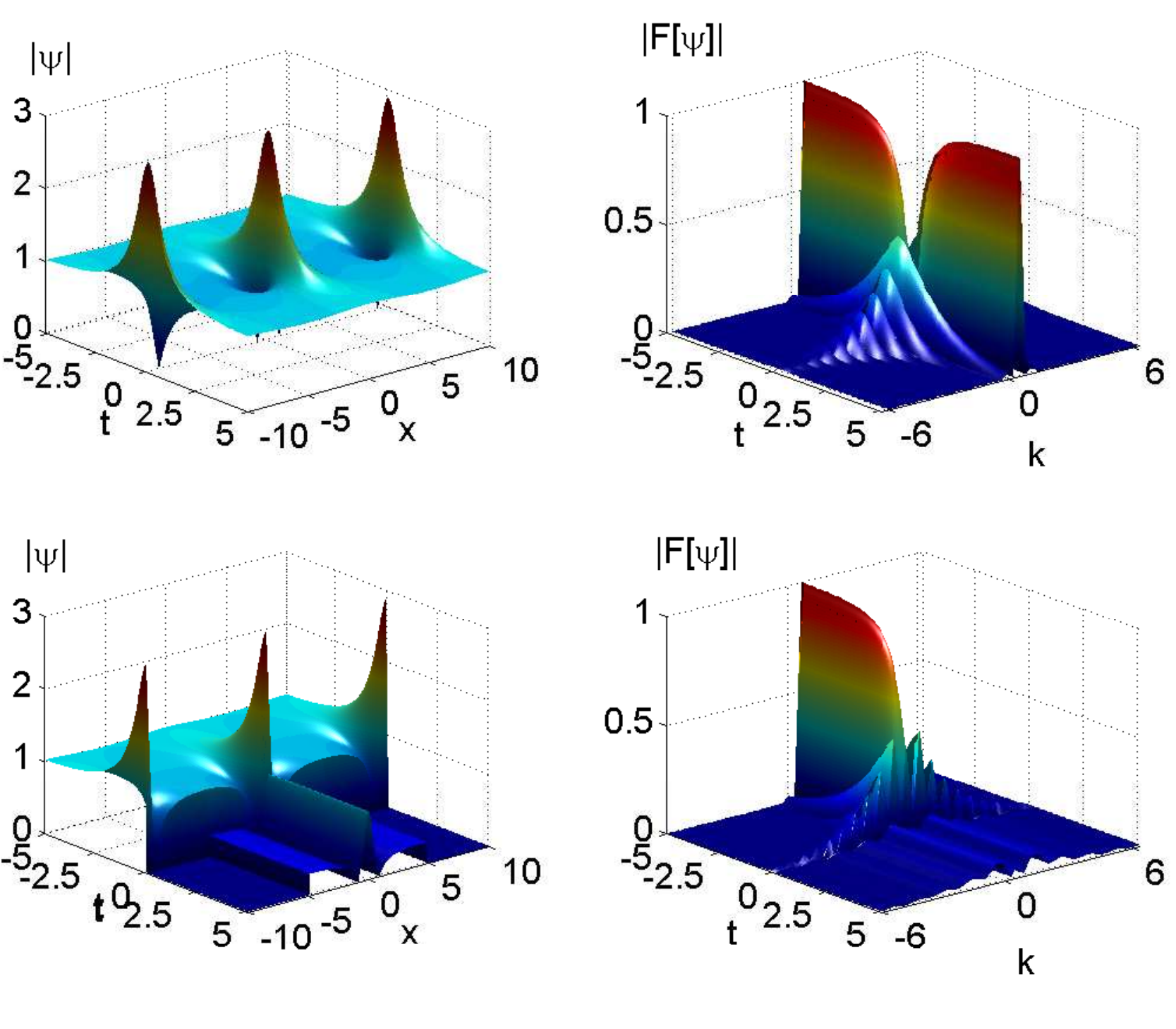}
  \end{center}
\caption{\small Zeno dynamics of an Akhmediev breather with $a=0.45$ a) freely evolving wave b) its Fourier spectrum c) continuously observed Akhmediev breather in $[-5,5]$ during $t=[0-5]$  d) its Fourier spectrum.}
  \label{fig2}
\end{figure}

Next we apply the similar procedure to the AB, but with a narrower region of observation which is selected as $L=[-5,5]$ to illustrate the effects of narrowing the observation domain as depicted in Fig.~\ref{fig2}. Due to Fourier duality principle, the AB frozen in this fashion has a wider spectrum compared to its counterpart presented in Fig.~\ref{fig1}.

Depending on the width of the Zeno observation domain, it is possible to freeze the whole AB and thus its triangular spectrum during the observation time.

\subsection{Freezing Peregrine Soliton}
Rogue waves of the NLSE is considered to be in the form of rational soliton solutions \cite{Akhmediev2009b}. The simplest rational soliton solution of the NLSE is the Peregrine soliton \cite{Peregrine, Kibler}. It is given by
\begin{equation}
\psi_1=\left[1-4\frac{1+2it}{1+4x^2+4t^2}  \right] \exp{[it]}
\label{eq13}
\end{equation}
where $t$ and $x$ denotes the time and space, respectively \cite{Akhmediev2009b}. This solution can be recovered as the limiting case of the AB when the period of the solution tends to infinity. It has been shown that Peregrine soliton is only a first order rational soliton solution of the NLSE \cite{Akhmediev2009b}. Higher order rational soliton solutions of the NLSE and a hierarchy of obtaining those rational solitons based on Darboux transformations are given in \cite{Akhmediev2009b}. Throughout many simulations \cite{Akhmediev2009b, Akhmediev2009a, Akhmediev2011} and some experiments \cite{Kibler}, it has been confirmed that rogue waves can be in the form of the first (Peregrine) and higher order rational soliton solutions of the NLSE.

\begin{figure}[htb!]
\begin{center}
   \includegraphics[width=3.4in]{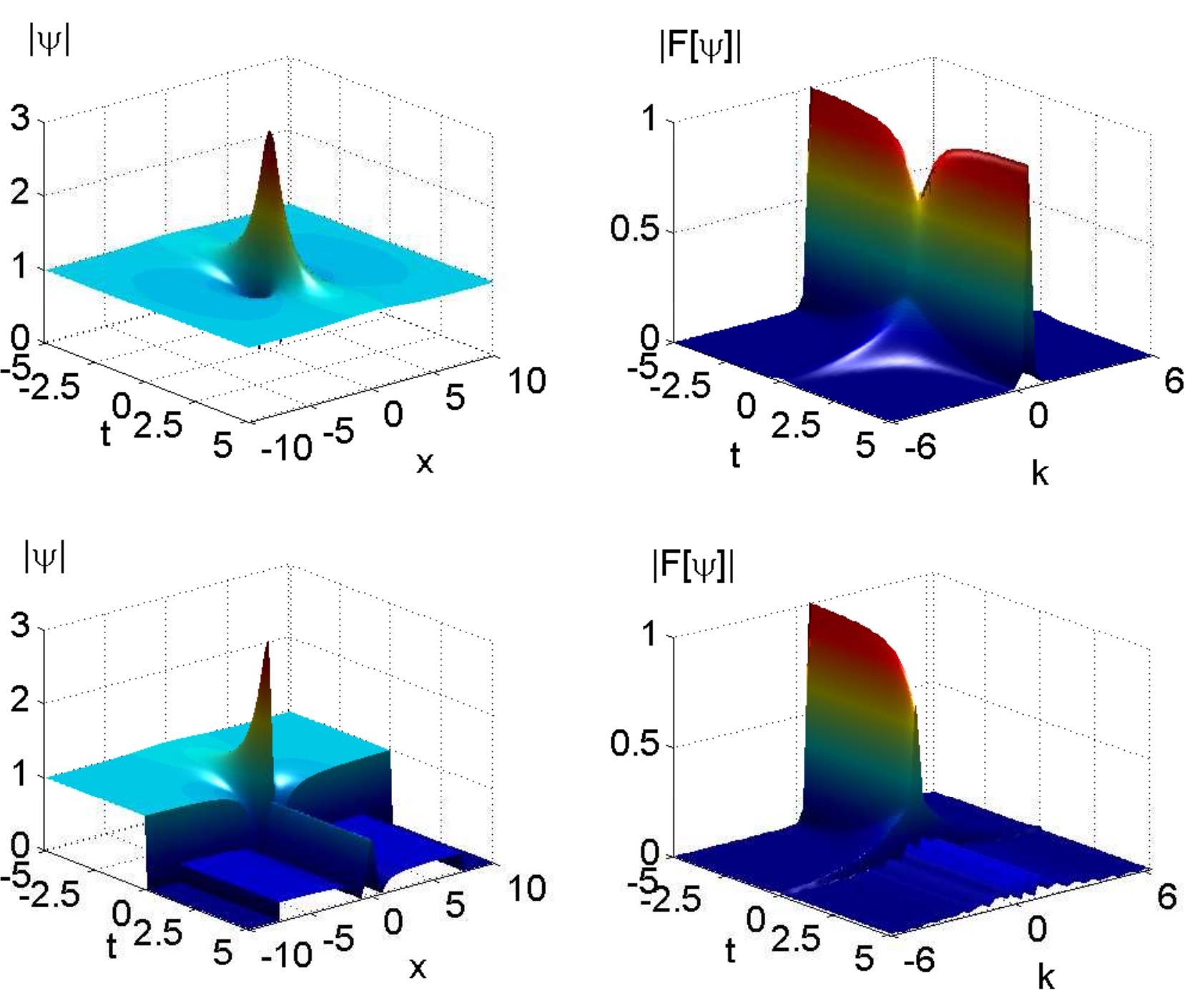}
  \end{center}
\caption{\small Zeno dynamics of a Peregrine soliton a) freely evolving wave b) its Fourier spectrum c) continuously observed Peregrine soliton in the interval of $[-7.5,7.5]$ during $t=[0-5]$  d) its Fourier spectrum.}
  \label{fig3}
\end{figure}

Similar to the AB case, in order to analyze the Zeno dynamics of the Peregrine soliton we apply the procedure described by Eqs.(\ref{eq07})-(\ref{eq09}). In Fig.~\ref{fig3}, plots of the Peregrine soliton and its Fourier spectrum are shown in the first row. This Peregrine soliton evolved freely during the time interval of $t=[-5,5]$. In the second row of Fig.~\ref{fig3}, we present the Peregrine soliton inhibited by Zeno observations and its corresponding Fourier spectrum. In this plot the Peregrine soliton evolved freely during the temporal interval of $t=[-5,0]$ and it was continuously subjected to Zeno observations in the time interval of $t=[0,5]$ within $L=[-7.5,7.5]$. For a better visualization of the effect of Zeno observations, the wave profile is not normalized in this figure as before. Again depending on the width of the Zeno observation domain, it is possible to freeze the Peregrine soliton wholly or partially and thus its triangular spectrum during the observation time.

\begin{figure}[htb!]
\begin{center}
   \includegraphics[width=3.4in]{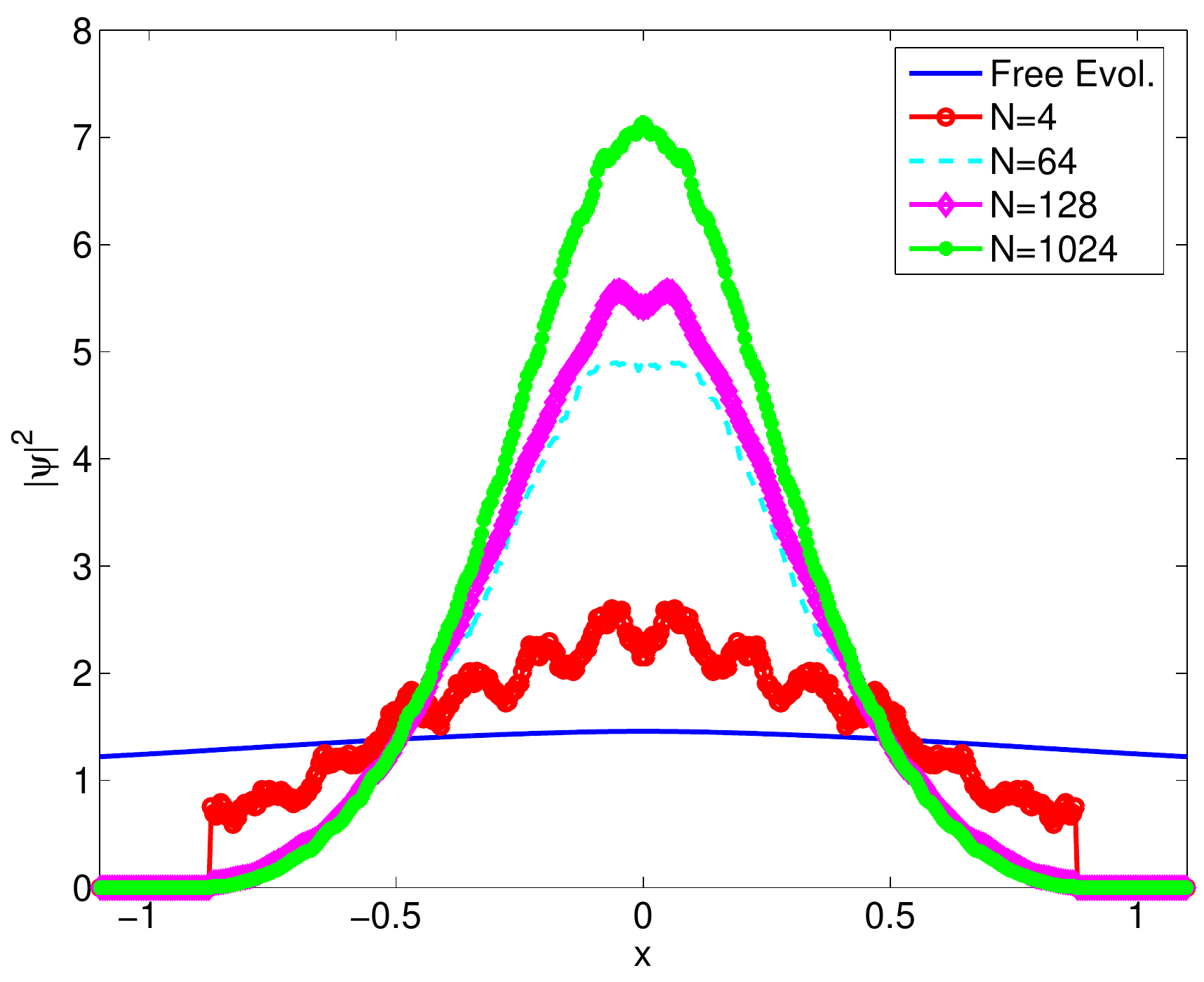}
  \end{center}
\caption{\small Probability density (unnormalized) after N intermediate measurements of the Peregrine soliton in the interval $[-0.8,0.8]$ for a time $t=n/N=2$.}
  \label{fig4}
\end{figure}

\begin{figure}[htb!]
\begin{center}
   \includegraphics[width=3.4in]{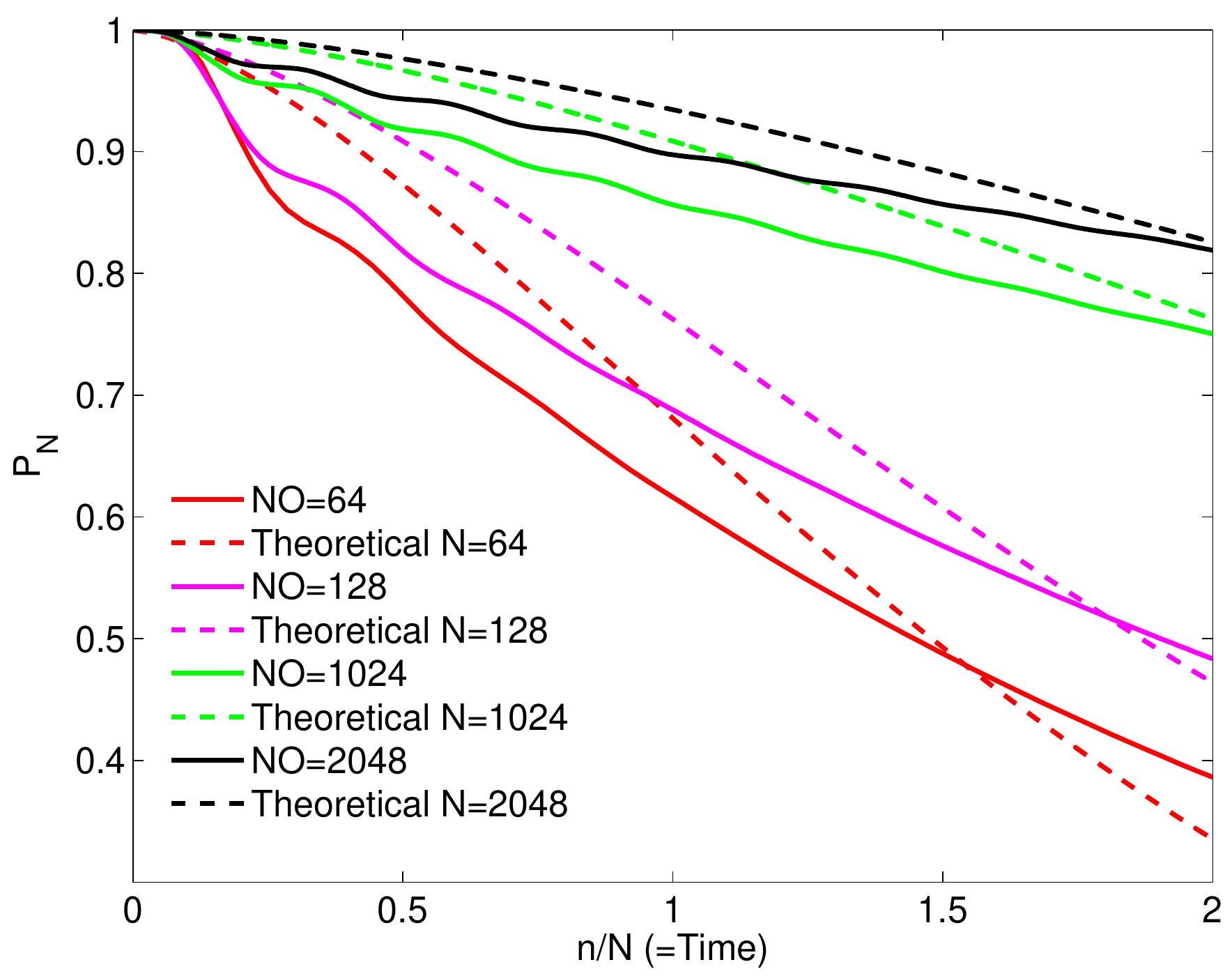}
  \end{center}
\caption{\small The probability of finding the Peregrine soliton in the interval $[-0.8,0.8]$ for a time of $t=n/N=2$ after N intermediate measurements.}
  \label{fig5}
\end{figure}

In Fig.~\ref{fig4}, we present the unnormalized wave profile with different observation numbers ($N$). For a larger number of intermediate measurements, the decay of Peregrine soliton is inhibited for longer times compared to the case of fewer Zeno observations. This is due to the fact that for a smaller number of observations the evolution time between two successive observations increases. Thus the wave profile diffuses in this relatively large temporal interval. The constant background level of unity is diminishes since the probability vanishes in the unobserved regions of the evolution domain for a positive measurement.

In Fig.~\ref{fig5}, we present the normalized probabilities of finding the wave in the observation domain for a time interval of $t=[0, 2]$. The dashed lines show the probabilities obtained analytically by Eqs.(\ref{eq10})-(\ref{eq11}) whereas the continuous lines represent numerical results. Since the numerical results are obtained using the NLSE and the analytical distributions given by Eqs.(\ref{eq10})-(\ref{eq11}) rely on the assumption that the particle's motion is governed by linear Schr\"{o}dinger equation, some discrepancies appear between two results, where the discrepancies are less for more frequent observations as expected.

\subsection{Freezing Akhmediev-Peregrine Soliton}

Second order rational soliton solution of the NLSE is Akhmediev-Peregrine soliton \cite{Akhmediev2009b}, which is considered to be a model for rogue waves with higher amplitude than the Peregrine soliton. The formula of Akhmediev-Peregrine soliton is given as
\begin{equation}
\psi_2=\left[1+\frac{G_2+it H_2}{D_2}  \right] \exp{[it]}
\label{eq14}
\end{equation}
where
\begin{equation}
G_2=\frac{3}{8}-3x^2-2x^4-9t^2-10t^4-12x^2t^2
\label{eq14}
\end{equation}
\begin{equation}
H_2=\frac{15}{4}+6x^2-4x^4-2t^2-4t^4-8x^2t^2
\label{eq16}
\end{equation}
and
\begin{equation}
\begin{split}
D_2=\frac{1}{8} & [ \frac{3}{4}+9x^2+4x^4+\frac{16}{3}x^6+33t^2+36t^4 \\
& +\frac{16}{3}t^6-24x^2t^2+16x^4t^2+16x^2t^4 ]
\label{eq17}
\end{split}
\end{equation}
where $t$ is the time and $x$ is the space parameter \cite{Akhmediev2009b}. Using Darboux transformation formalism this soliton can be obtained using the Peregrine soliton as the seed solution \cite{Akhmediev2009b}. Many numerical simulations also confirm that rogue waves in the NLSE framework can also be in the form of Akhmediev-Peregrine soliton \cite{Akhmediev2011, Akhmediev2009b, Akhmediev2009a}. However, to our best knowledge an experimental verification of this soliton still does not exist.

\begin{figure}[htb!]
\begin{center}
   \includegraphics[width=3.4in]{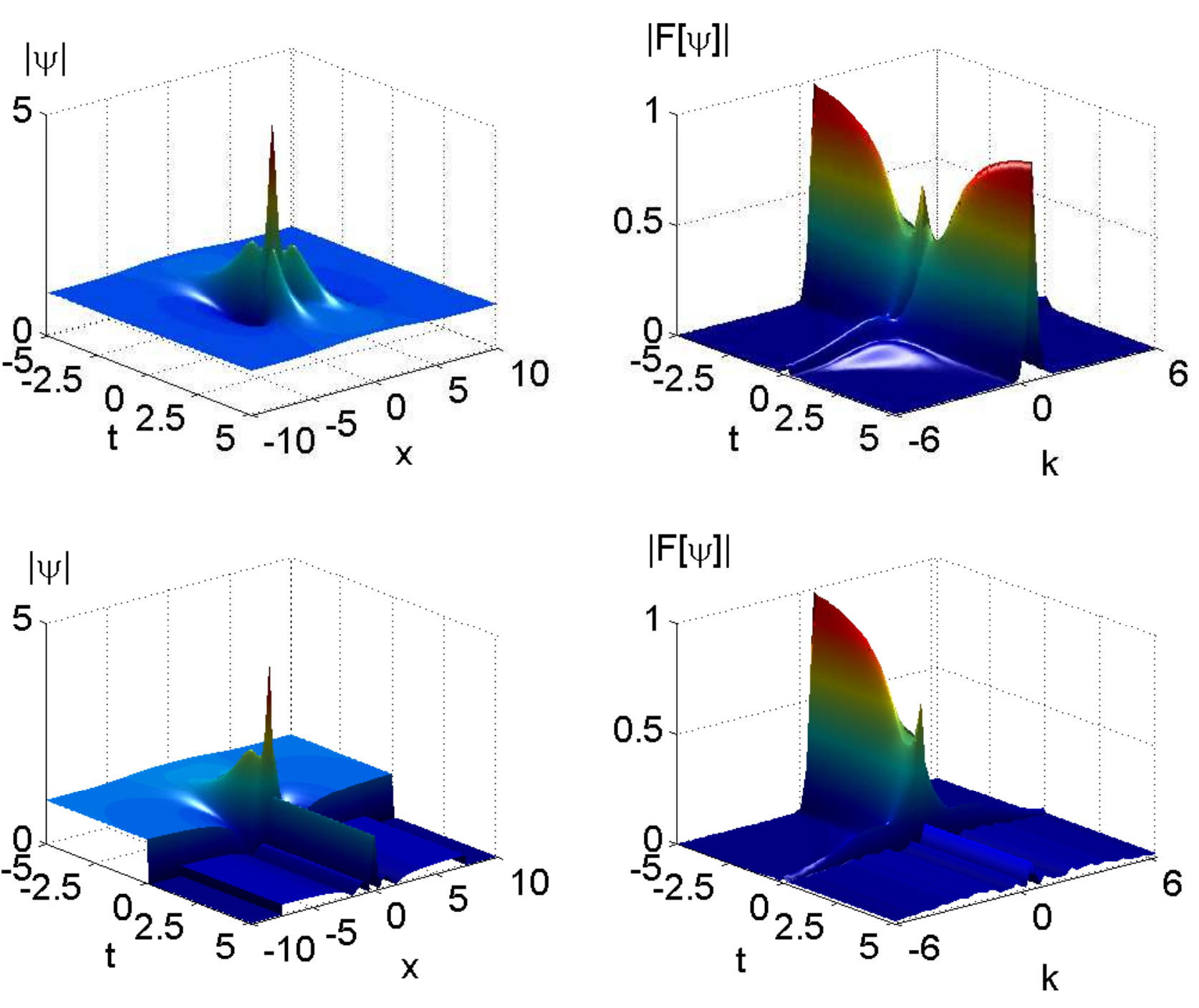}
  \end{center}
\caption{\small Zeno dynamics of an Akhmediev-Peregrine soliton a) freely evolving wave b) its Fourier spectrum c) continuously observed Akhmediev-Peregrine soliton in the interval of $[-7.5,7.5]$ during $t=[0-5]$  d) its Fourier spectrum.}
  \label{fig6}
\end{figure}

As in the AB and Peregrine soliton cases, in order to analyze the Zeno dynamics of the Akhmediev-Peregrine soliton we apply the procedure described by Eqs.(\ref{eq07})-(\ref{eq09}). Similarly, in Fig.~\ref{fig6}, the Akhmediev-Peregrine soliton, which evolved freely during the time interval of $t=[-5,5]$, and its Fourier spectrum are plotted in the first row. In the second row of Fig.~\ref{fig6}, the Akhmediev-Peregrine soliton inhibited by Zeno observations and its corresponding Fourier spectrum are depicted. In this plot the Akhmediev-Peregrine soliton evolved freely during the temporal interval of $t=[-5,0]$ and continuous Zeno observations took place time within $L=[-7.5,7.5]$ during the time interval of $t=[0,5]$.

\begin{figure}[htb!]
\begin{center}
   \includegraphics[width=3.4in]{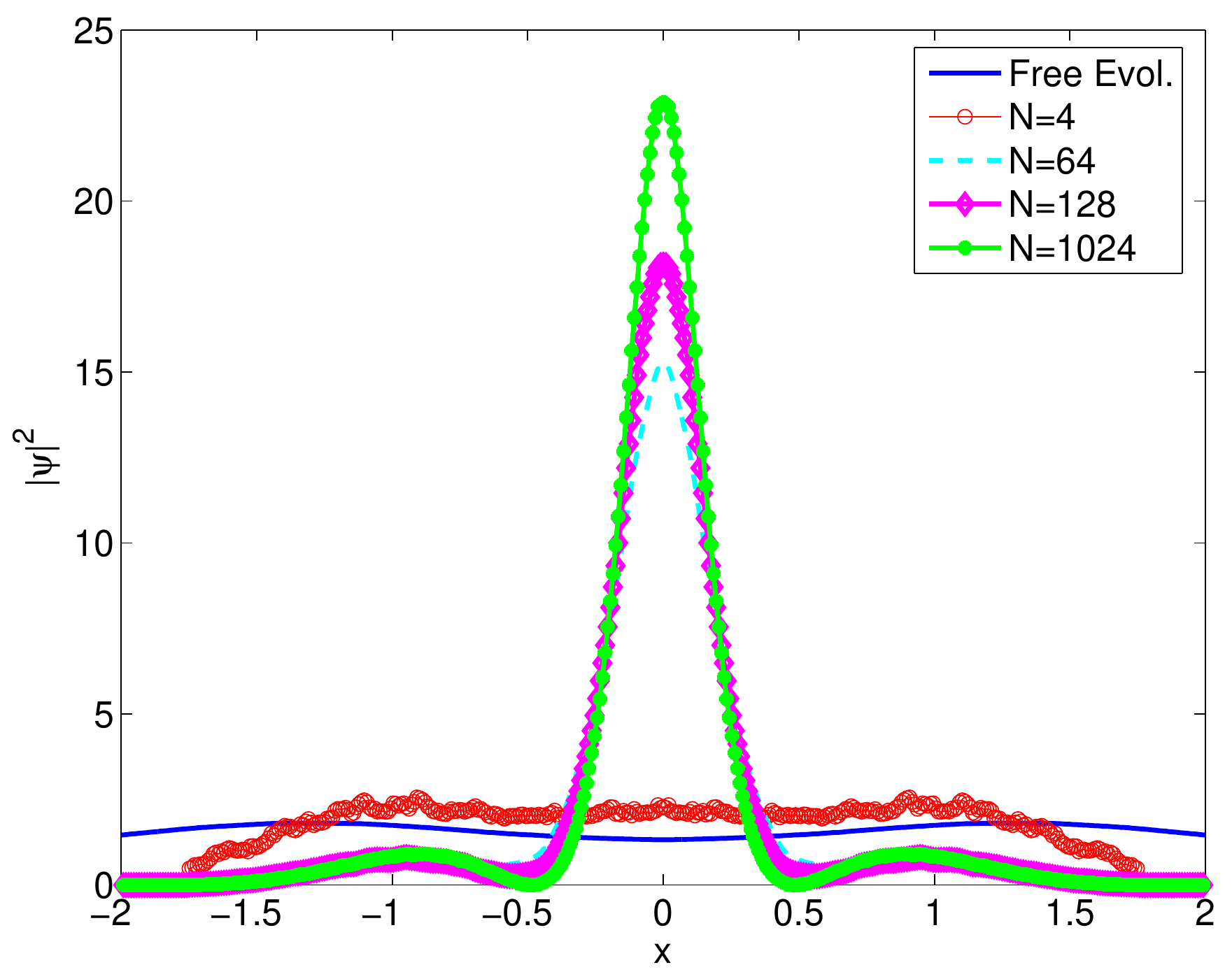}
  \end{center}
\caption{\small Probability density after N intermediate measurements of the Akhmediev-Peregrine soliton in the interval $[-1.7,1.7]$ for a time $t=n/N=2$.}
  \label{fig7}
\end{figure}
Again unnormalized wave profile is depicted in this figure for a better visualization of the effect of Zeno observations on the Akhmediev-Peregrine soliton.  Again depending on the width of the Zeno observation domain, it is possible to freeze the Akhmediev-Peregrine soliton wholly or partially and thus its triangular spectrum can be preserved during the observation time.

In Fig.~\ref{fig7}, the unnormalized wave profiles for different observation numbers ($N$) are shown. Similar to the previous cases, for a larger number of intermediate measurements, the decay of Akhmediev-Peregrine soliton is inhibited for longer times compared to the case of fewer Zeno observations due to shorter diffusion time between two positive Zeno observations.  Additionally, the peak as well as two dips of the Akhmediev-Peregrine soliton can be preserved during measurements depending on the length of the observation domain. Similarly, since the probability vanishes in the unobserved regions of the evolution domain for a positive measurement, the constant background level of unity is diminishes.

\begin{figure}[h!]
\begin{center}
   \includegraphics[width=3.4in]{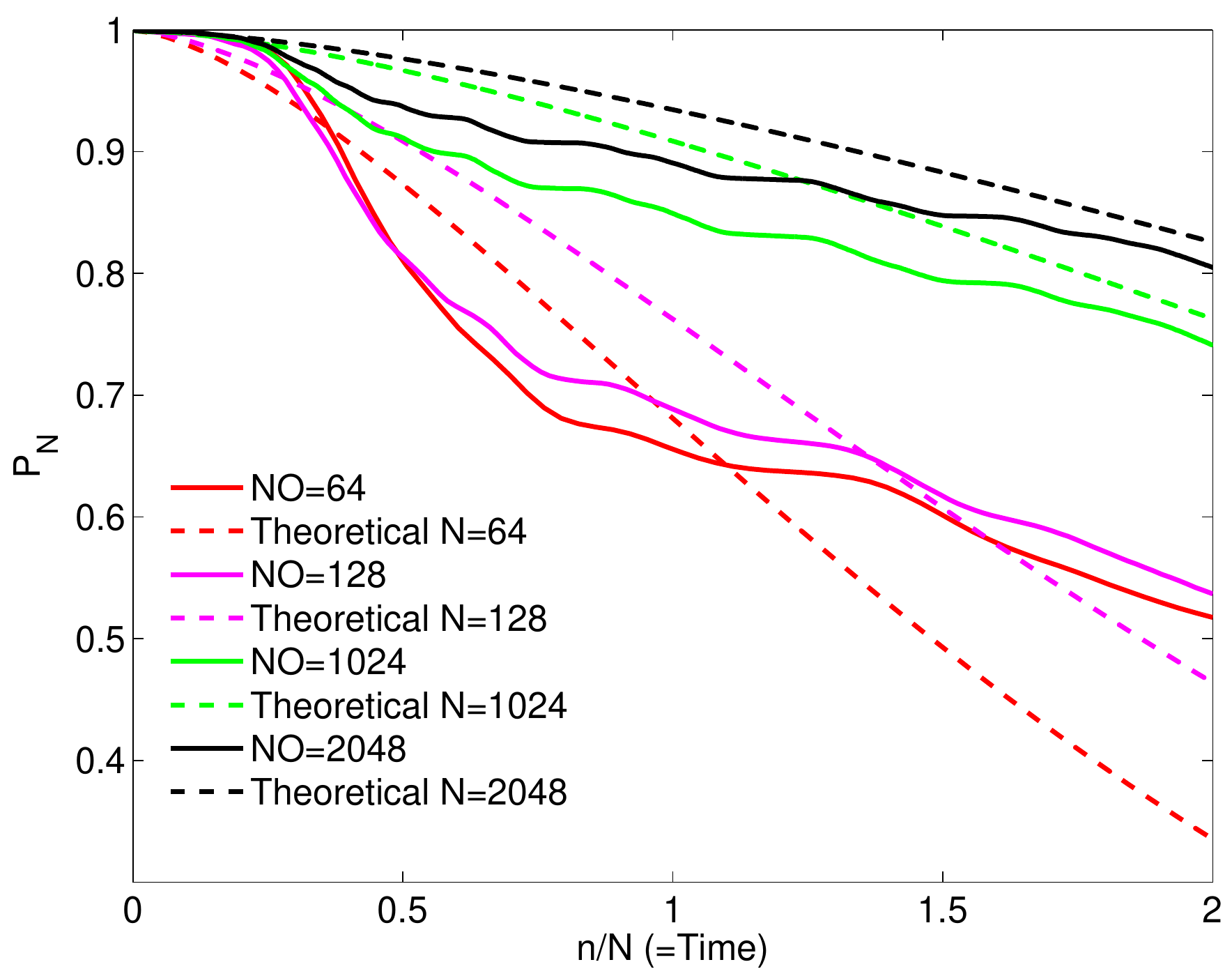}
  \end{center}
\caption{\small The probability of finding the Akhmediev-Peregrine soliton in the interval $[-0.8,0.8]$ for a time of $t=n/N=2$ after N intermediate measurements.}
  \label{fig8}
\end{figure}
In Fig.~\ref{fig8}, we present the normalized probabilities of finding the wave in the observation domain for a time interval of $t=[0, 2]$. The dashed lines show the probabilities obtained analytically by Eqs.(\ref{eq10})-(\ref{eq11}) whereas the continuous lines represent numerical results. As before, since the numerical results are obtained using the NLSE and the analytical distributions given by Eqs.(\ref{eq10})-(\ref{eq11}) rely on the assumption that the particle's motion is governed by linear Schr\"{o}dinger equation, some discrepancies appear between two results, where the discrepancies are less for more frequent observations as expected. However the discrepancies are slightly higher compared to the Zeno dynamics of the Peregrine soliton depicted in Fig.~\ref{fig5} due to increased steepness of the Akhmediev-Peregrine soliton.

\section{Conclusion}

In this paper we have numerically investigated the Zeno dynamics of optical rogue waves in the frame of the standard NLSE. In particular, we have analyzed the Zeno dynamics of the Akhmediev breathers, Peregrine and Akhmediev-Peregrine soliton solutions of the NLSE, which are considered as accurate rogue wave models. We have showed that frequent measurements of the rogue wave inhibits its movement in the observation domain for each of these solutions. We have analyzed the spectra of the rogue waves to observe the supercontinuum generation under Zeno observations. Fourier as well as the wavelet spectra of rogue waves under Zeno dynamics may give some clue about the application time (position) of the Zeno observations to freeze the emergence or decay of rogue waves. This would especially be important for the Zeno dynamics of stochastic wavefields which produce rogue waves. We have also analyzed the effect of observation frequency on the rogue wave profile and on the probability of freezing the wave in the observation domain. We have showed that the rogue wave shape can be preserved for longer times and the probability of the freezing the rogue wave increases for more frequent observations, for all three types of rogue waves considered.

The revival dynamics, that is the dynamics after Zeno observations are ceased, of rogue waves will be a part of future work. The analysis of statistical distributions and the shapes of rogue waves in the stochastic fields after Zeno observation remains as an problem which needs further attention as well. 
We believe that the results presented herein may open new insights in modelling the dynamics of standing and propagating optical rogue waves. 
More specifically, the procedure analyzed in this paper in the frame of the standard NLSE can be used to advance the fields of optical communications and vibrations with special applications which include but are not limited to freezing and steering the rogue and other wave types, avoiding their breaking, imposing a time delay by Zeno effect. The procedure analyzed in this paper in the frame of the standard NLSE can also be extended to model the Zeno dynamics of the many other fascinating nonlinear phenomena which can be used to advance the optical science and technology.

\section*{Acknowledgment}
FO is funded by Isik University Scientific Research Funding Agency under Grant Number: BAP-15B103.


\end{document}